\documentstyle[12pt,epsfig,twoside]{article}
\newcommand{\bei}{\begin{itemize}}
\newcommand{\eei}{\end{itemize}}
\newcommand{\beq}{\begin{equation}}
\newcommand{\eeq}{\end{equation}}
\newcommand{\beqn}{\begin{eqnarray}}
\newcommand{\eeqn}{\end{eqnarray}}
\newcommand{\beqns}{\begin{eqnarray*}}
\newcommand{\eeqns}{\end{eqnarray*}}

\begin{document}


\today

\vspace{-0.2cm}

\begin{center} 
{\bf {\large Comments on "An Update of the HLS Estimate of the Muon g-2" by M.~Benayoun {\it et al.}, arXiv:1210.7184v3} \\}
\vspace{1.0cm}

M.~Davier$^{\,\mathrm a}$\footnote
{ 
	davier@lal.in2p3.fr
}
,
B.~Malaescu$^{\,\mathrm b}$\footnote
{
	malaescu@in2p3.fr
} \\

\vspace{0.5cm}
{\small \em $^{\mathrm a}$Laboratoire de l'Acc\'el\'erateur Lin\'eaire,\\
IN2P3-CNRS et Universit\'e de Paris-Sud, 91898 Orsay, France}\\
\vspace{0.1cm}
{\small \em $^{\mathrm b}$Laboratoire de Physique Nucl\'eaire et des Hautes Energies, IN2P3-CNRS et Universit\'es Pierre-et-Marie-Curie et Denis-Diderot, 75252 Paris Cedex 05, France}\\
\vspace{1.0cm}

{\small{\bf Abstract}}
\end{center}
{\small
\vspace{-0.2cm}
In a recent paper~\cite{benayoun} M.~Benayoun {\it et al.} use a specific 
model to compare results on the existing data for the cross section of the 
process $e^+e^-\rightarrow \pi^+\pi^-$ and state conclusions about the 
inconsistency of the BABAR results with those from the other experiments.
We show that a direct model-independent comparison of the data at hand
contradicts this claim. Clear discrepancies with the results of 
Ref.~\cite{benayoun} are pointed out. As a consequence we do not 
believe that the lower value and the smaller uncertainty obtained for the 
prediction of the muon magnetic anomaly are reliable results.   
\noindent
}
\vspace{25mm}



%
%
\section{Introduction}
\label{sec_introduction}

The authors of Ref.~\cite{benayoun} use a specific model of the cross section
for the process $e^+e^-\rightarrow \pi^+\pi^-$ in order to compute the
corresponding contribution to the muon magnetic anomaly. Specifically they 
fix parameters of a theoretical model based on the concept of Hidden Local 
Symmetry (references to be found in Ref.~\cite{benayoun}) using the spectral 
function from $\tau$ decays obtained by the Belle collaboration~\cite{belle} 
and other less precise inputs from additional processes assuming vector dominance. 

In these comments we consider four points relevant to the analysis in 
Ref.~\cite{benayoun} where the model is used to discriminate between the 
different sets of recent $e^+e^-$ data, allowing the authors to make 
statements about their consistency. 

First, their model for the $e^+e^-$ cross section relies on the $\tau$
spectral function from Belle within the framework of the HLS approach. 
Such a procedure relies on assumptions regarding the isospin-breaking 
corrections needed in order to link the $\tau$ and $e^+e^-$ spectral functions.
We provide some comments on this point in Sec.~\ref{IB}.
 
Second, the model is used by the authors to compare the results of the model 
with existing measurements of the $\pi^+\pi^-$ annihilation process. In the 
peak $\rho$ region they claim good agreement between the data from 
KLOE~\cite{kloe-08,kloe-10}, CMD-2~\cite{cmd2-03,cmd2-06}, and SND~\cite{snd},
while pointing out a discrepancy with BABAR~\cite{babar-prl,babar-prd}. 
As we show in Sec.~\ref{rho-peak} this conclusion is not consistent with 
the direct comparison of these data.

Third, still using the model, a specific analysis of the $\rho-\omega$ 
interference pattern in the above experimental data is performed, again 
showing an even bigger discrepancy between BABAR and the other experiments.
However the $\rho-\omega$ interference has been already extracted by the 
experiments (but KLOE) and the published results clearly do not support the 
claim made by the authors, as demonstrated in Sec.~\ref{rho-omega}.
 
Fourth, we comment on the values obtained for the hadronic contribution to
the muon $g-2$ and on the stated increased significance of the discrepancy 
with the direct measurement~\cite{bnl} which we believe is not warranted.

In the following we address these four issues. For the second and third 
points, we show that in a model-independent way the direct comparison of 
the measured cross sections yields results which are at variance with those 
of Ref.~\cite{benayoun}.

\section{Isospin Breaking Corrections}
\label{IB}

In addition to the standard isospin-breaking (IB) 
corrections~\cite{adh,cirigliano,dehz03,castro,tauee}, 
M.~Benayoun {\it et al.} consider the
effect of $\gamma-\rho$ mixing, already proposed previously~\cite{jeger}. 
A major consequence is a mass shift between neutral and charged $\rho$ 
resonances. But such an effect was already considered in the analysis of 
Ref.~\cite{tauee}, together with electromagnetic contributions in the $\rho$ 
decays, with the conclusion that the isospin-corrected $\tau$ spectral function
agreed well with the BABAR data and rather well with the Novosibirsk results,
while disagreeing with KLOE. 

The authors of Ref.~\cite{benayoun} reach the opposite conclusion, {\it i.e.}
good agreement with KLOE, fair agreement with CMD-2 and SND, and discrepancy 
with BABAR. We point out that even if the corrections used follow the same 
rules as in Ref.~\cite{jeger}, their final results differ significantly. 
Therefore it seems that the IB corrections as
used by the authors are still subject to some debate, especially since they
do not provide specific studies of their systematic uncertainties.
In these conditions it is probably not safe to use the IB-corrected $\tau$ 
spectral function as an exact replica of the $e^+e^-$ cross section when 
comparing to the $e^+e^-$ data.

\section{Consistency of Different Data in the Peak $\rho$ Region}
\label{rho-peak}

The direct comparison of the cross sections for BABAR and the other experiments
has been shown in detail in Ref.~\cite{babar-prd}. There is indeed a 
discrepancy with KLOE, but agreement with CMD-2 and SND within the quoted 
respective systematic uncertainties. One can quantify the observations by 
considering the ratio of the cross sections measured by each experiment to the 
BABAR result. Since the energy corresponding to the experimental data points 
do not match the BABAR bins, we perform the quantitative comparison to the 
form factor fit of the BABAR data, described in Ref.~\cite{babar-prd} which 
nicely interpolates the data.
 
\begin{figure}[htp] 
\centering
   \includegraphics[width=7.5cm]{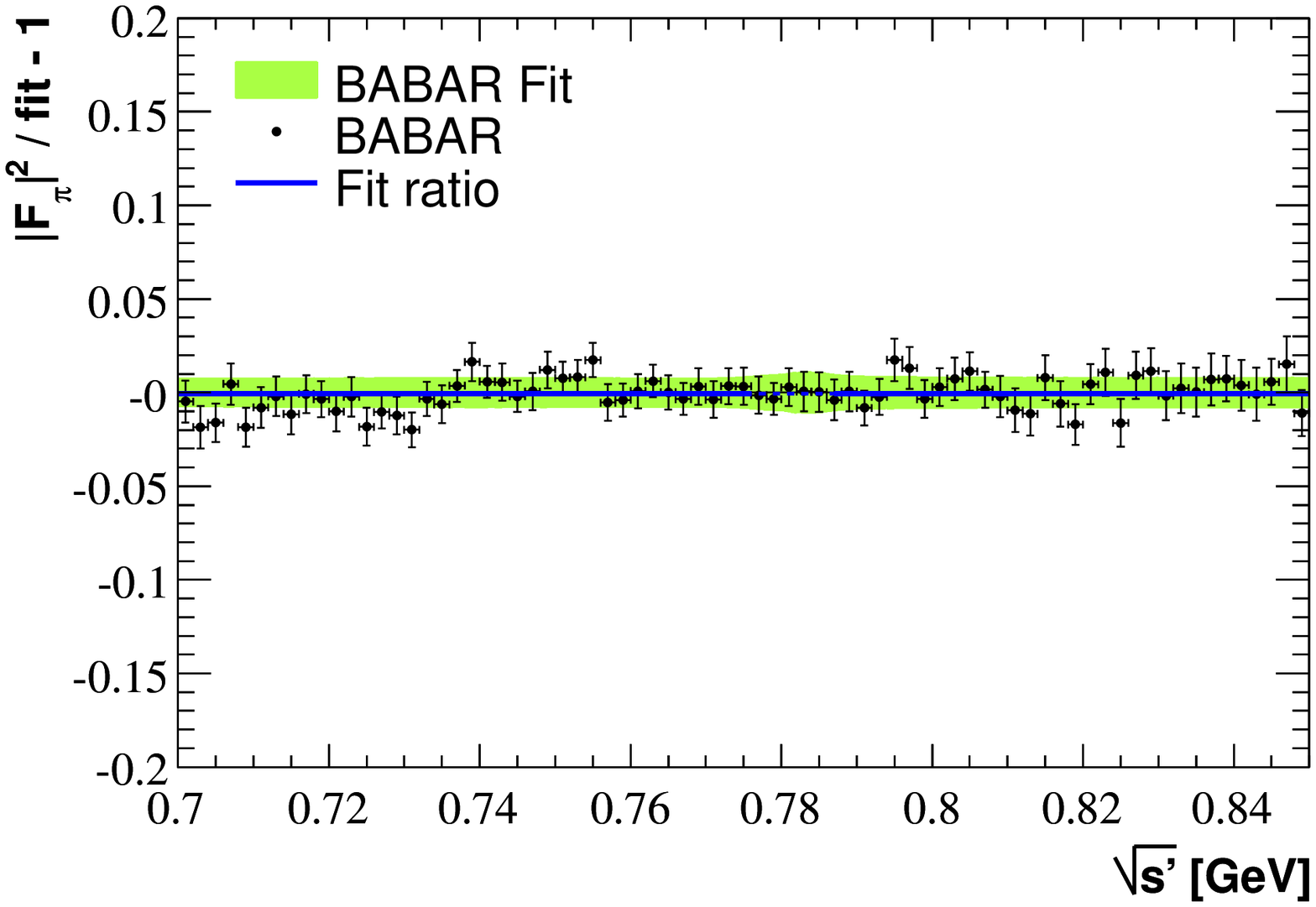}
   \includegraphics[width=7.5cm]{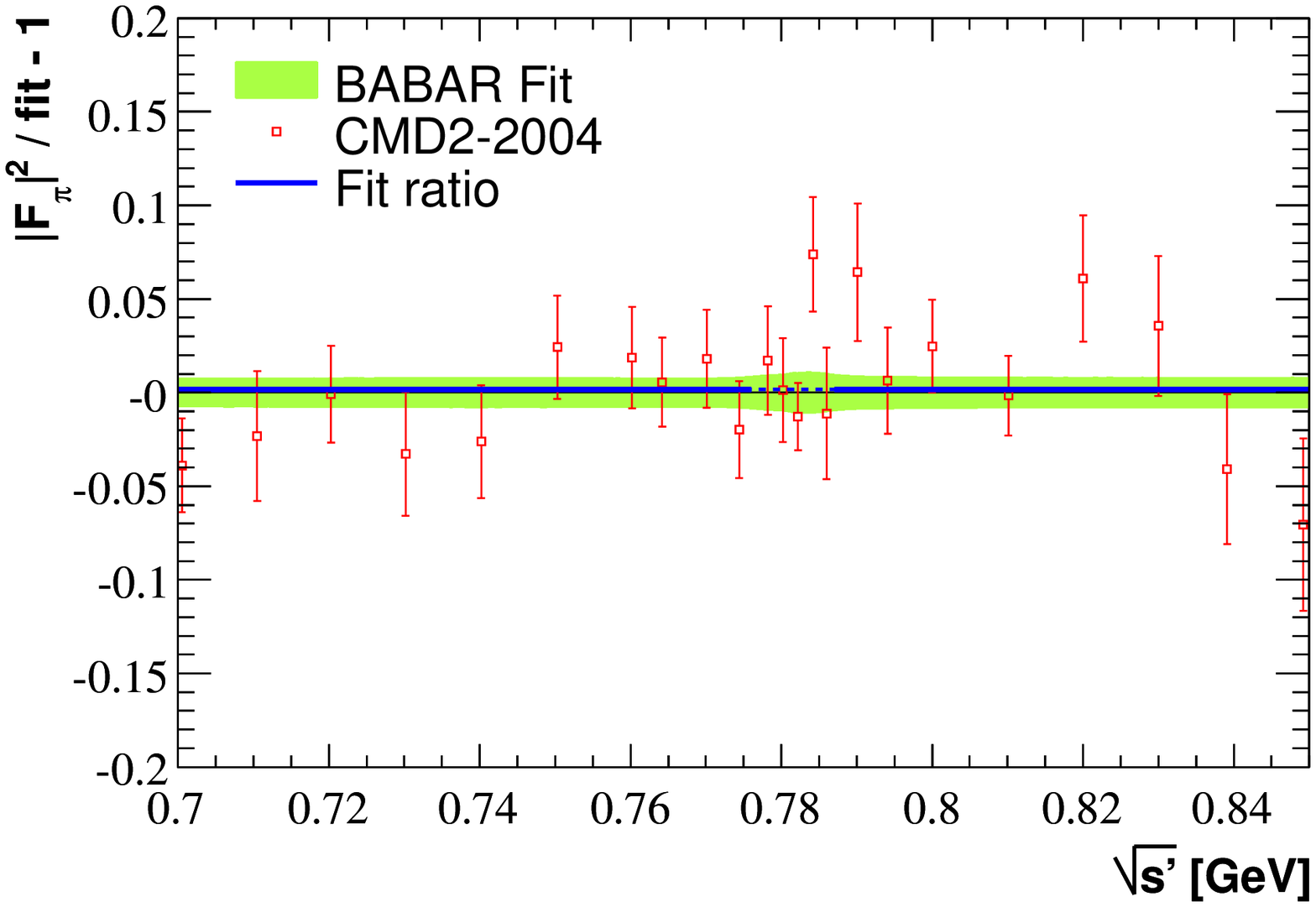} \\
   \includegraphics[width=7.5cm]{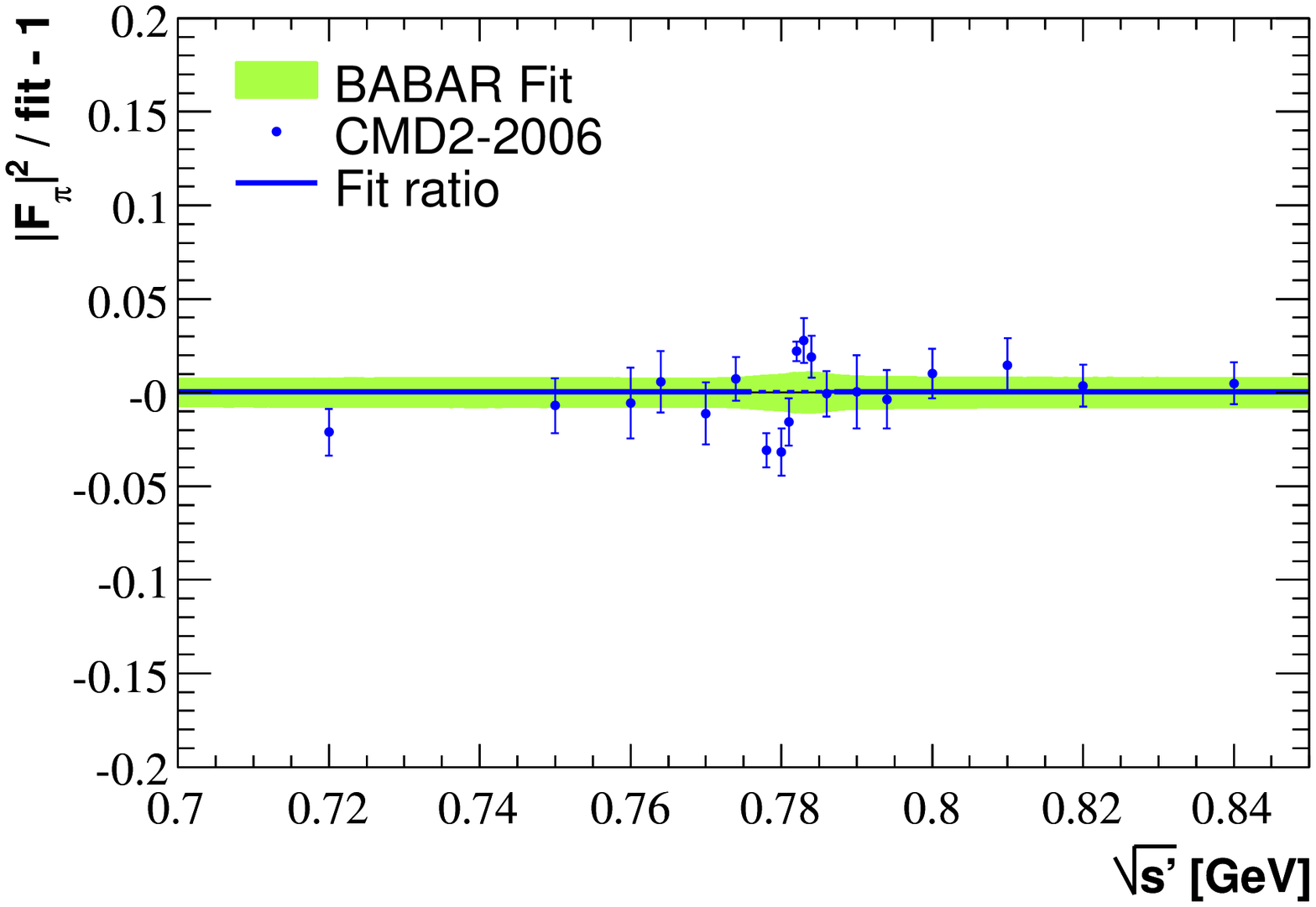}
   \includegraphics[width=7.5cm]{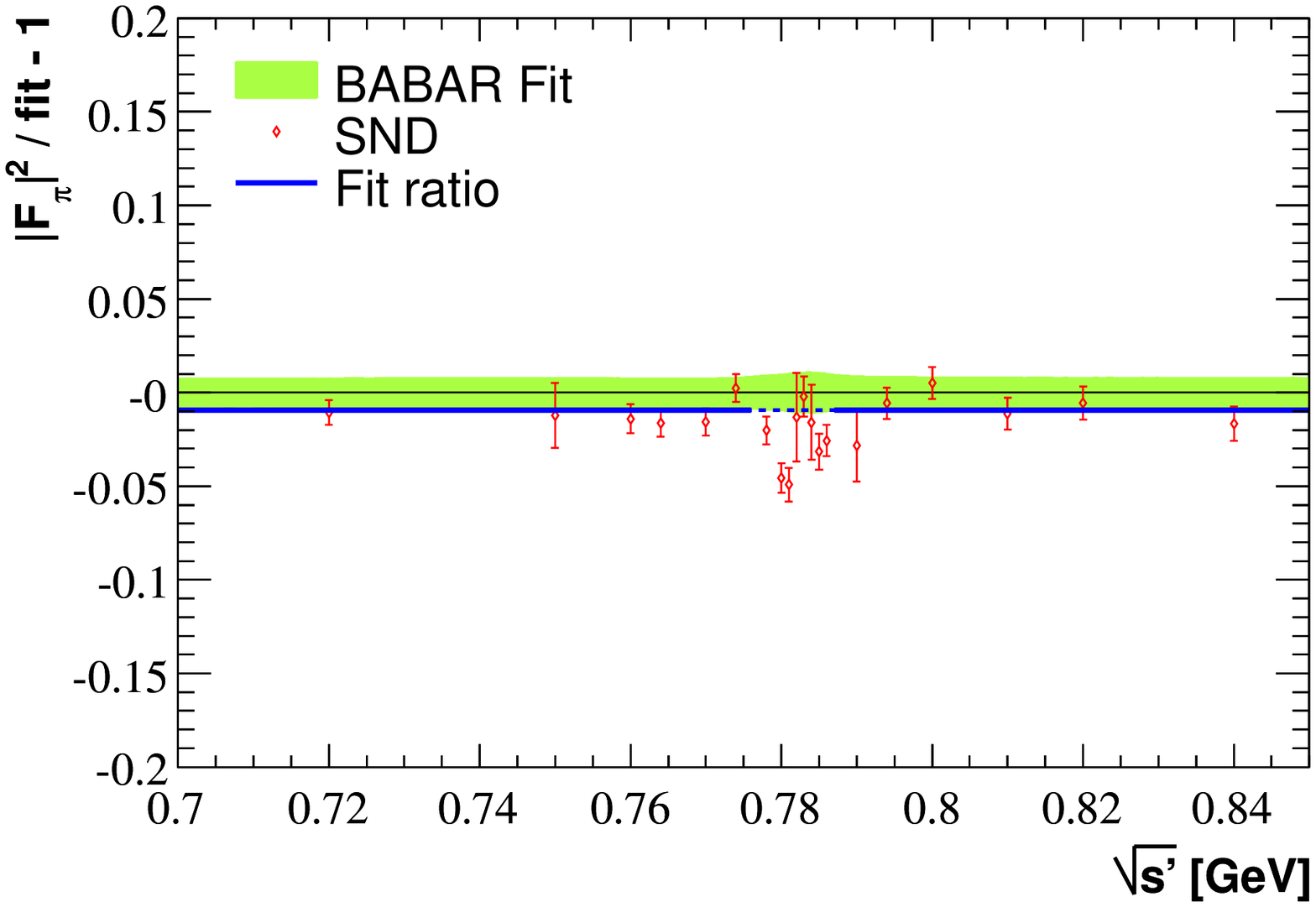} \\
   \includegraphics[width=7.5cm]{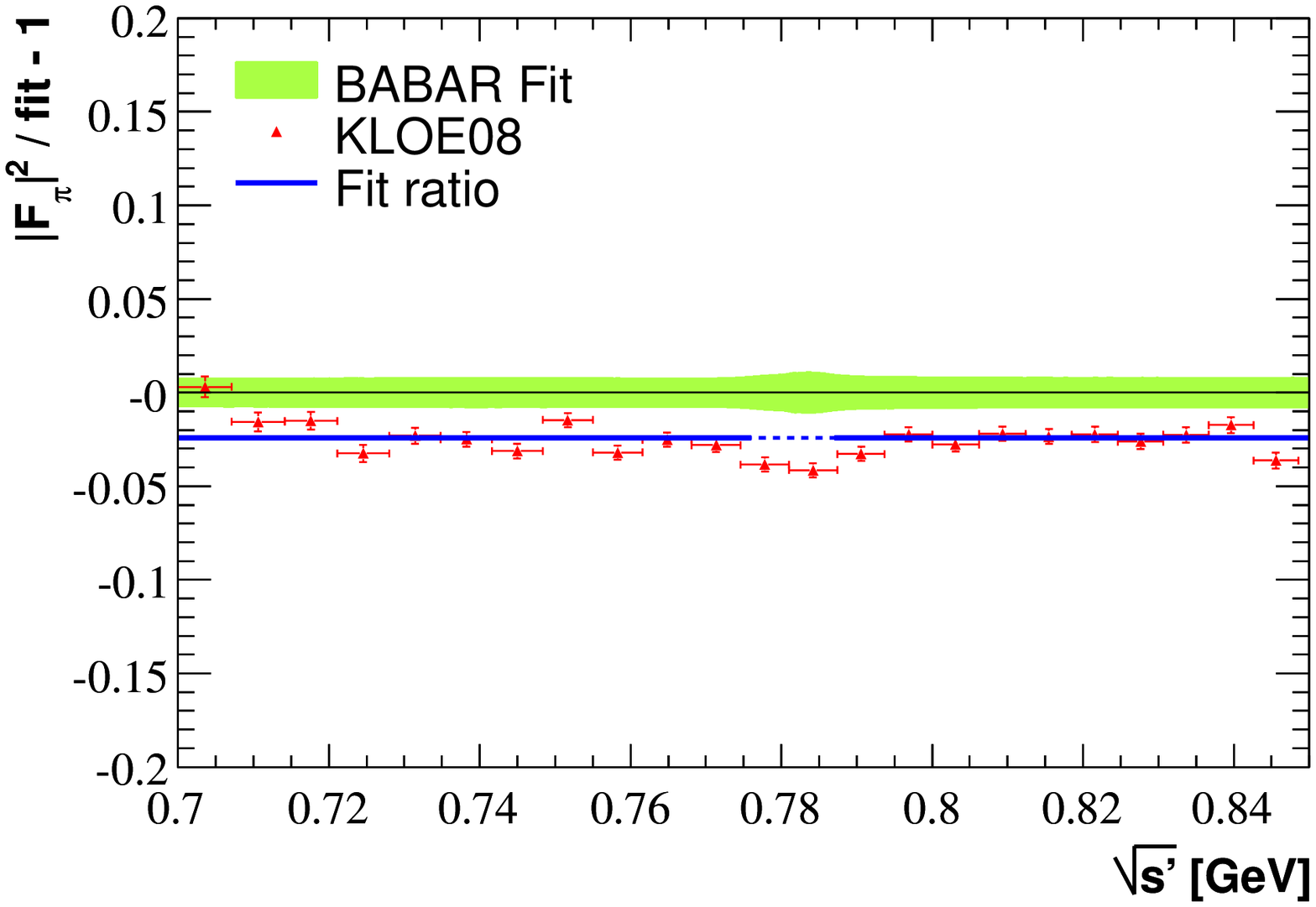}
   \includegraphics[width=7.5cm]{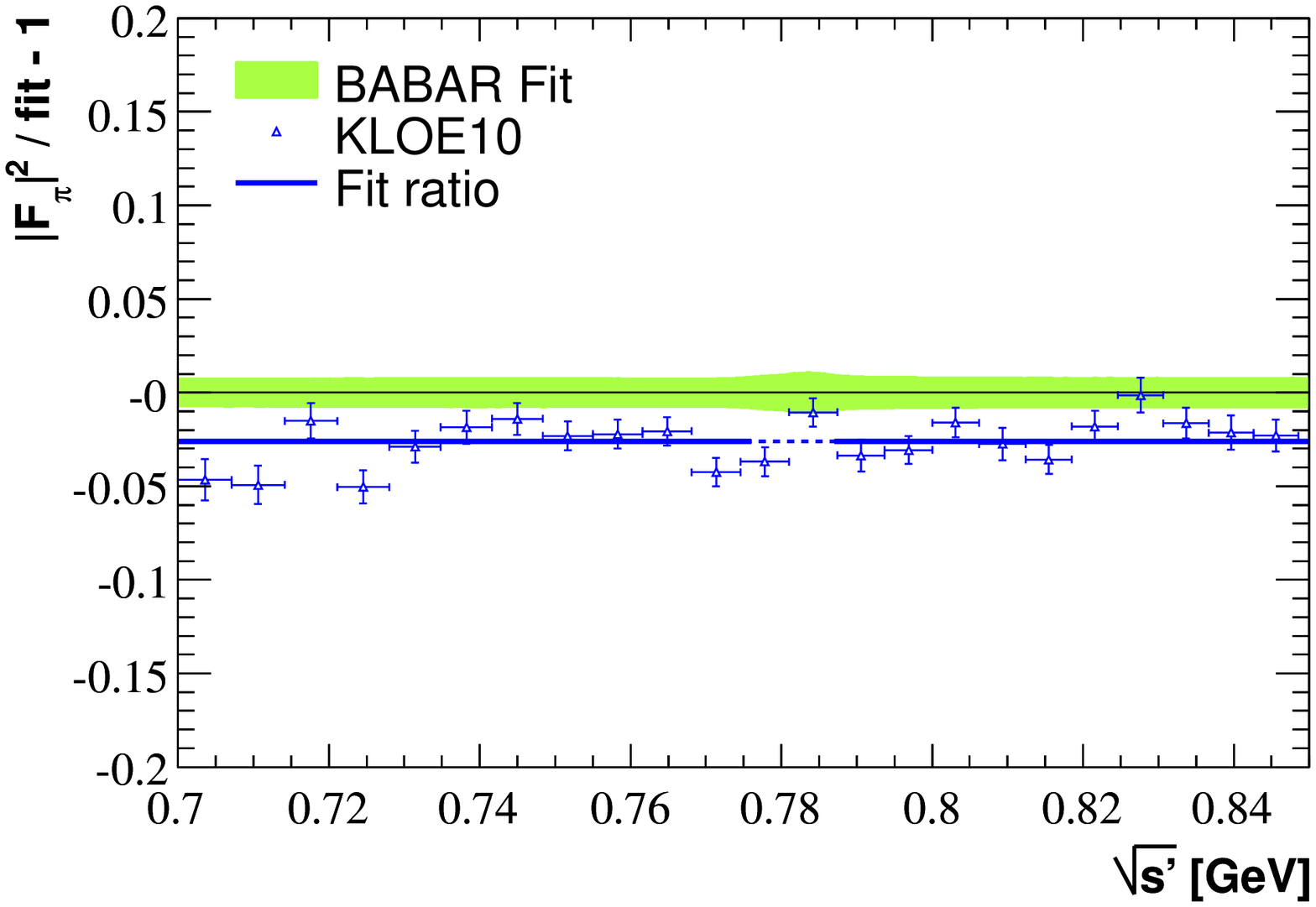}
  \caption{\label{fit-ratios} \small  Fits to a constant of the ratios 
(minus 1) of the 
measured $e^+e^-\rightarrow \pi^+\pi^-$ cross sections from the quoted 
experiments relative to the BABAR cross section fit in the mass range 0.7-0.85 
GeV, excluding the $\rho-\omega$ interference region. Only statistical 
uncertainties are used in the fits and plotted.The result of the fits is shown
with a blue line, while the green band around 0 gives the region allowed by 
BABAR statistical and systematic uncertainties.} 
\end{figure}

From inspection of their Fig.5 the authors of Ref.~\cite{benayoun} point out 
major differences in the $\rho$ peak region 0.70-0.85 GeV. Thus we restrict 
the ratio fits to this mass range. We exclude the $\rho-\omega$ region 
0.776-0.788 GeV, discussed in the next section, as the ratios are very 
sensitive to the relative mass calibration of the experiments (this effect 
is not taken into account in Ref~\cite{benayoun}).
The ratios to BABAR as a function of $\pi\pi$ mass are displayed in 
Fig.~\ref{fit-ratios}. The fit to a constant is a good description of the 
respective data, except for KLOE where deviations are seen at lower masses, 
however not affecting significantly the fit over most of the mass range.

\begin{figure}[htp] 
\centering
  \includegraphics[width=8cm]{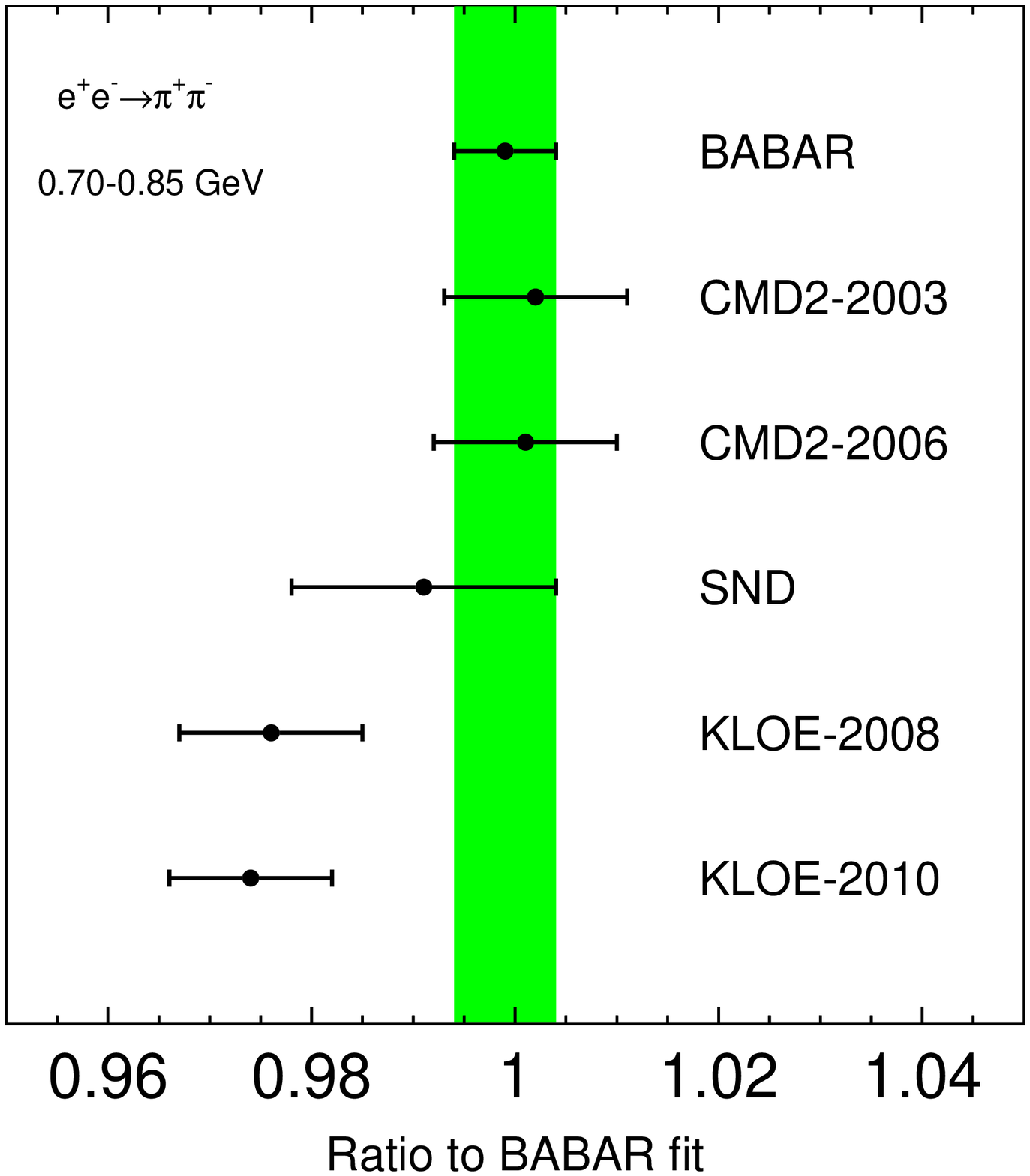}
  \caption{\label{ratios} \small  Ratios of the measured 
$e^+e^-\rightarrow \pi^+\pi^-$ cross sections from the quoted experiments 
relative to the BABAR cross section fit in the mass range 0.7-0.85 GeV, 
excluding the $\rho-\omega$ interference region. The quoted error bars include
both statistical and systematic uncertainties.} 
\end{figure}

The fitted values for the ratios are given in Table~\ref{table1}. Of course 
the BABAR value in the first row is very close to one by construction, 
but we show it to demonstrate that the form factor fit, performed in the full 
mass range from theshold to 3 GeV, does represent well the BABAR data between 
0.7 and 0.85 GeV. Only statistical uncertainties are used in the ratio fits, 
while systematic uncertainties are quoted separately. All the values are 
displayed in Fig.~\ref{ratios}.

\begin{table}\centering
\label{table1}
\caption{\small Ratios of the measured 
$e^+e^-\rightarrow \pi^+\pi^-$ cross sections from the quoted experiments 
relative to the BABAR form factor fit in the mass range 0.7-0.85 GeV, excluding the $\rho-\omega$ interference region . The quoted uncertainties are 
statistical ('stat'), systematic ('syst'), and total ('tot'). The $\chi^2$  
per degree of freedom ($DF$) of the ratio fits to a constant are given.}
\vspace{0.5cm}
\begin{tabular}{|c|c|c|c|c|c|} \hline\hline
Experiment/BABAR-fit & ratio & 'stat' & $\chi^2/DF$ & 'syst' & 'tot'\\ \hline
 BABAR      & 0.999 & 0.001 & 52.5/60 & 0.005 & 0.005 \\
 CMD-2 2003 & 1.002 & 0.007 & 18.6/17 & 0.006 & 0.009 \\
 CMD-2 2006 & 1.001 & 0.004 &  6.1/11 & 0.008 & 0.009 \\
 SND        & 0.991 & 0.002 &  9.4/11 & 0.013 & 0.013 \\
 KLOE 2008  & 0.976 & 0.001 & 66.8/20 & 0.009 & 0.009 \\
 KLOE 2010  & 0.974 & 0.002 & 39.8/20 & 0.008 & 0.008 \\
\hline\hline
\end{tabular}
\end{table}

From Table~\ref{table1} and Fig~\ref{ratios} we see that, unlike the 
qualitative claim in Ref.~\cite{benayoun}, the BABAR data in the $\rho$ peak 
region is fully compatible with CMD-2 and SND, and disagrees with KLOE.
The CMD-2 data also disagrees with KLOE, while SND with a larger uncertainty 
is consistent with both BABAR and KLOE. Since in Ref.~\cite{benayoun} CMD-2 
and SND are treated together, it is interesting to quote the ratio for the
combined Novosibisrk data (taking into account correlated uncertainties). Its
value of $0.999\pm0.007$ shows the same disagreement with KLOE as BABAR. 
Therefore we conclude that the claim by Benayoun {\it et al.} is incorrect.
There is no justified argument for discarding the BABAR data in the 
evaluation of the $g-2$ dispersion integral.

\section{Analysis of $\rho-\omega$ Interference}
\label{rho-omega}

Except for KLOE, all the experiments have provided fits of the pion form factor
with $\rho$ and $\omega$ amplitudes, even including isovector resonances at
higher mass. In these analyses the parameters describing the $\rho$ resonance 
and the interference amplitude are fitted, so that the product of the branching
ratios $B_{\omega\rightarrow ee}\cdot B_{\omega\rightarrow\pi\pi}$ can be 
directly obtained. Using the fit results published by the experiments, the 
values deduced for the product are given in Table~\ref{table2}. The values
presented in Ref.~\cite{benayoun} are listed for comparison. The agreement 
is reasonable for CMD-2 and SND, however there is a large discrepancy for
BABAR. Whereas the value from the published BABAR fit to the pion form factor 
agrees well with Novosibirsk, the value from Ref.~\cite{benayoun} is
problematic: it deviates considerably from the direct value and the quoted 
uncertainty is one order of magnitude too small, given the statistical accuracy
of the BABAR data. Thus the conclusion reached in Ref.~\cite{benayoun} that the
BABAR data strongly disagree with the Novosibirsk experiments is ill-founded.
 
\begin{table}
\centering
\label{table2}
\caption{\small The results for the product
$B_{\omega\rightarrow ee}\cdot B_{\omega\rightarrow\pi\pi}$ obtained from
pion form factor fits published by the experiments are compared to the
results presented in Ref.~\cite{benayoun}. The conclusions regarding the BABAR
value and its uncertainty are radically different in the two cases.}
\vspace{0.5cm}
\begin{tabular}{|c|c|c|} \hline\hline
Experiment  & $B_{ee}^\omega\cdot B_{\pi\pi}^\omega~(10^{-6})$ [exp] & $B_{ee}^\omega\cdot B_{\pi\pi}^\omega~(10^{-6})$ Ref.~\cite{benayoun}\\ \hline
 CMD-2 2003 & $0.95 \pm 0.18$ & - \\
 CMD-2 2006 & $1.02 \pm 0.09$ & - \\
 SND        & $1.22 \pm 0.07$ & - \\
 CMD-2 + SND& $1.13 \pm 0.05$ & $1.22 \pm 0.04$ \\
 BABAR      & $1.05 \pm 0.08$ & $1.78 \pm 0.01$ \\
\hline\hline
\end{tabular}
\end{table}

\section{The muon magnetic anomaly}

The authors of Ref.~\cite{benayoun} present a range of values indicating a 
discrepancy with the direct measurement~~\cite{bnl}, with their 'best' estimate
given at 4.9$\sigma$. This result follows from two facts: the prediction has 
a lower central value, as the result of discarding the BABAR data, and its 
uncertainty is reduced because their model brings additional constraints.

In comparison the most recent analyses~\cite{dhmz2010,hlmnt2011} are 
model-independent and use all the available $e^+e^-$ data, taking into account
the existing discrepancies between experiments in order to set realistic 
uncertainties. In that case the discrepancy in the muon magnetic anomaly is
only at 3.3-3.7$\sigma$.

\section{Conclusions}

We have pointed out serious discrepancies in the model-dependent analysis of
Ref.~\cite{benayoun} when compared to the original
$e^+e^-\rightarrow \pi^+\pi^-$ cross section data. Let us summarize our points:
\begin{itemize}
\item The approach relies crucially on model-dependent isospin-breaking 
corrections involved in relating the $\tau$ spectral function to the $e^+e^-$ 
cross section. A detailed study of the corresponding systematic uncertainties 
is lacking.
\item The direct and quantitative comparison of the different $e^+e^-$ data 
sets contradicts the claim made that BABAR data disagree with the other 
experiments.
\item The direct BABAR analysis of $\rho-\omega$ interference gives a result
in agreement with Novosibirsk data, in contradiction with the value given in 
Ref.~\cite{benayoun} which makes no sense and is again used to discriminate 
against the BABAR data. 
\item As a consequence we do not believe that the lower value and the smaller 
uncertainty obtained for the prediction of the muon magnetic anomaly are 
reliable results.
\end{itemize}

%
%

{\small
\begin{thebibliography}{99}
\bibitem{benayoun}  M.~Benayoun {\it et al.}, arXiv:1210.7184v3.
\bibitem{belle}    M. Fujikawa {\it et al.}, 
                      Phys. Rev. {\bf D 78}, 072006 (2008).
\bibitem{kloe-08}  F. Ambrosino {\it et al.},
                         Phys. Lett. {\bf B670}, 285 (2009).
\bibitem{kloe-10}   F. Ambrosino {\it et al.},
                           Phys. Lett. {\bf B700}, 102-110 (2011).
\bibitem{cmd2-03}    R.R. Akhmetshin {\it et al.},
                           Phys. Lett. {\bf B578}, 285 (2004).
\bibitem{cmd2-06}    R.R. Akhmetshin {\it et al.},
                           Phys. Lett. {\bf B648}, 28 (2007).
\bibitem{snd} M.N. Achasov {\it et al.},
                JETP {\bf 103}, 380 (2006).
\bibitem{babar-prl}  B.~Aubert {\it et al.},
                           Phys. Rev. Lett. {\bf 103}, 231801 (2009).
\bibitem{babar-prd}  J.P.~Lees {\it et al.},
                           Phys. Rev. {\bf D86}, 032013 (2012).
\bibitem{bnl}         G.W. Bennett {\it et al.},
                           Phys. Rev. {\bf D73}, 072003 (2006).
\bibitem{adh}     R.~Alemany, M.~Davier, and A.~Hocker,
                        Eur. Phys. J. {\bf C2} 123 (1998).
\bibitem{cirigliano}     V. Cirigliano, G. Ecker, and H. Neufeld, 
                           Phys. Lett. {\bf B513}, 361 (2001);
                           J. High Energy Phys. {\bf 08}, 002 (2002).
\bibitem{dehz03}      M. Davier, S. Eidelman, A. H\"ocker, and Z. Zhang,
                           Eur. Phys. J. {\bf C27}, 497 (2003);
                         {\bf C31}, 503 (2003). 
\bibitem{castro}       A. Flores-Talpa {\it et al.}, 
                           Phys. Rev. {\bf D74}, 071301 (2006);
                           Nucl. Phys. Proc. Suppl. {\bf 169}, 250 (2007).
\bibitem{tauee}   M. Davier {\it et al.},  
                           Eur. Phys. Jour. {\bf C66}, 127 (2010).
\bibitem{jeger}  F.~Jegerlehner and R.~Szafron,
                           Eur. Phys. J. {\bf C71}, 1632 (2011).
\bibitem{dhmz2010}     M.~Davier, A.~Hoecker, B.~Malaescu, and Z.~Zhang,
		           Eur. Phys. J. {\bf C 71}, 1515 (2011).
\bibitem{hlmnt2011}  K.~Hagiwara {\it et al.}, 
                           J. Phys. G 38, 085003~(2011).
\end{thebibliography} 
}
\end{document}